\documentclass[11pt,twoside]{article}
\usepackage{apn3conf}

\def\edcomment#1{\iffalse\marginpar{\raggedright\sl#1\/}\else\relax\fi}
\reversemarginpar

\begin{document}

\title{New Planetary Nebulae in the Northern Galactic Plane}
\titlemark{New PNe in the Northern Galactic Plane}

\author{Panayotis Boumis}
\affil{Institute of Astronomy \& Astrophysics, National Observatory of Athens, I. Metaxa \& V. Paulou, GR-152 36 P. Penteli, Athens, Greece, Email: ptb@astro.noa.gr}
\author{Myfanwy Bryce}
\affil{Jodrell Bank Observatory, Department of Physics \& Astronomy, University of Manchester, Macclesfield, Cheshire SK11 9DL, UK, Email: mbryce@ast.man.ac.uk}

\contact{Panayotis Boumis}
\email{ptb@astro.noa.gr}

\paindex{Boumis, P.}
\aindex{Bryce, M.}

\authormark{Boumis \& Bryce}

\keywords{ISM: planetary nebulae: general - Galaxy: bulge}

\begin{abstract}
We present deep H$\alpha +$[N {\sc ii}] CCD images of selected new
Planetary Nebulae (PNe) discovered through an [O {\sc iii}]
5007$\AA$~emission line survey in the Galactic Bulge region with $l >
0^{\rm o}$. In total, we detected more than 200 objects, including 45
new PNe. Deep H$\alpha +$[N {\sc ii}] CCD images as well as 
low resolution spectra were obtained for the new PNe in order to study
them in detail. Their first morphological study suggests that most of
them are round (60\%), while 31\% and 9\% are elliptical and bipolar,
respectively. 63\% of the round PNe present ring-like structure while
37\% are compact.
\end{abstract}

\section{Introduction}
Galactic Planetary Nebulae (PNe) are of great interest because of
their important role in the chemical enrichment history of the
interstellar medium as well as in the star formation history and
evolution of our Galaxy (Beaulieu et al. 2000 and references
therein). Many surveys have been made in the past in order to discover
new PNe. A list with all PNe catalogues can be found in Boumis et
al. (2003a - Paper I). The study of PNe offers the opportunity to
determine basic physical parameters (like morphology, kinematics,
abundances, distances, masses etc.), which will help to test
theoretical models and understand more about the dynamics of the
Galaxy. The morphology of PNe provide us the opportunity to have a
better picture of how the star evolve. Their morphology is a parameter
which is difficult to quantify and several studies have been made in
the past to produce a morphological classification scheme for PNe
(Balick 1987; Schwarz et al. 1993; Manchado et al. 2000 and references
therein). It is accepted that PNe having different morphologies arise
from progenitors with different masses. In particular, circular PNe
derive from stars with low mass progenitor, bipolar arise from higher
mass stars while elliptical derive from all masses of progenitor
(Phillips 2003 and references therein).  On a sample of PNe,
Stanghellini et al. (1993; 2002) found that the central star
distribution is different for bipolar and elliptical. In their latest
work, they also confirm that round, elliptical and bipolar PNe have
different spatial distributions within the Galaxy and that they might
belong to different stellar populations. Here, we present selected
H$\alpha+$[N {\sc ii}] images of new PNe showing the different
morphological types found, and our first results concerning their
morphology.

\section{Observations}

\subsection{The [O {\sc iii}] 5007 \AA\ Survey}
The survey observations were performed during the 2000--2001 observing
seasons, with the 0.3 m telescope at Skinakas Observatory in Crete
(Greece). We observed the regions $10^{o} < l < 20^{o}$, $-10^{o} < b
< -3^{o}$ and $0^{o} < l < 20^{o}$, $3^{o} < b < 10^{o}$ because (a)
the site of the Observatory allows us to observe down to $-25^{o}$
declination and (b) the [O {\sc iii}] 5007 \AA\ emission line is
absorbed between $-3^{o} < b < 3^{o}$. Our aim was to discover PNe
which (1) are characterized by strong [O {\sc iii}] 5007 \AA\
emission, (2) are slightly extended, because of our spatial
resolution, (3) are point--like with signal to noise greater than 4,
(4) do not have declination below -25$^{o}$ and (5) are observed only
during dark time periods to avoid the moonlight scattering effect on
the [O {\sc iii}] 5007 \AA\ line. The observational details and the
detection method are given in Paper I and Boumis \& Papamastorakis
(2003), respectively.

\subsection{Imaging \& Spectroscopy}
\begin{figure}
\centering
\mbox{\epsfclipon\epsfxsize=1.8in\epsfbox[20 20 575 575]{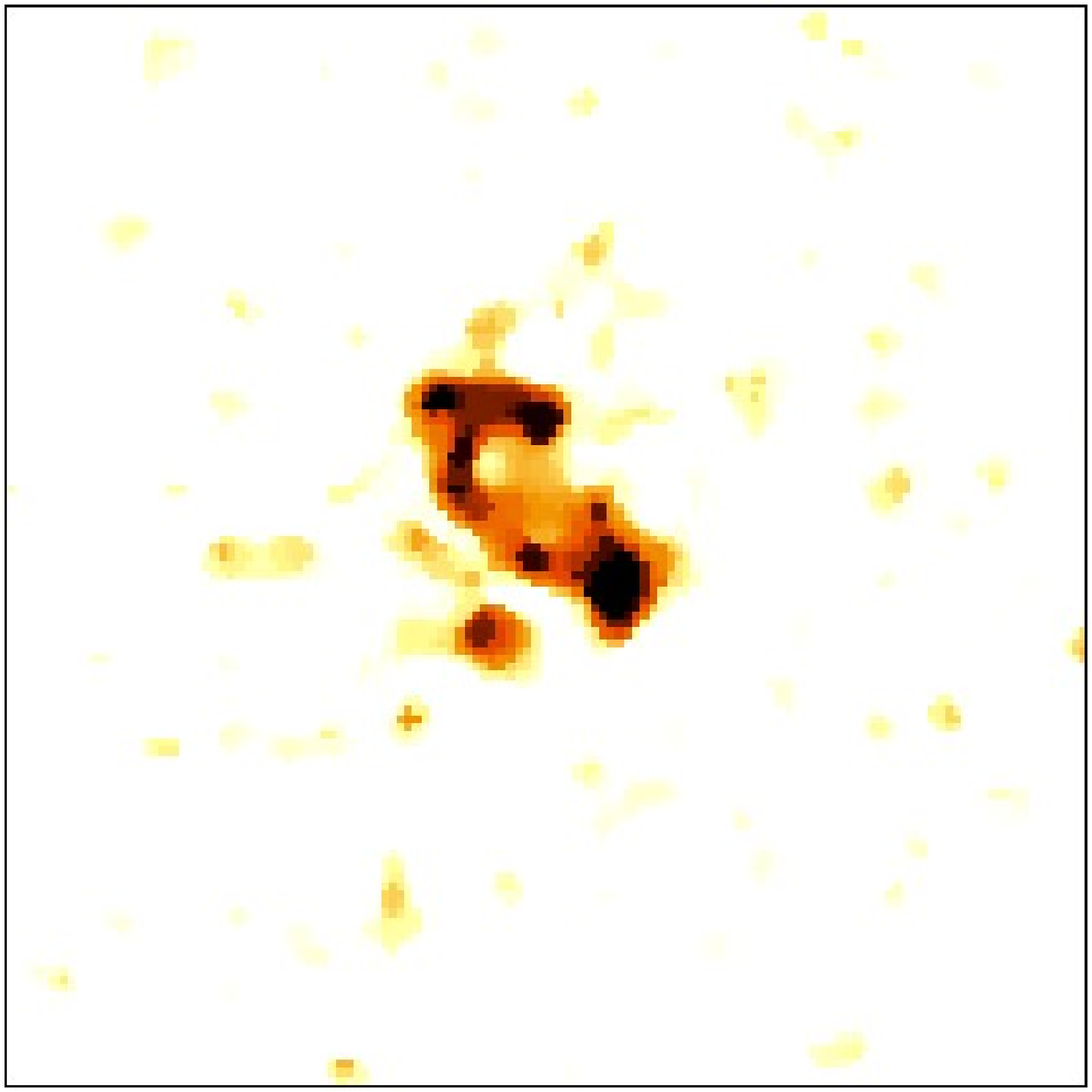}}
\mbox{\epsfclipon\epsfxsize=1.8in\epsfbox[20 20 575 575]{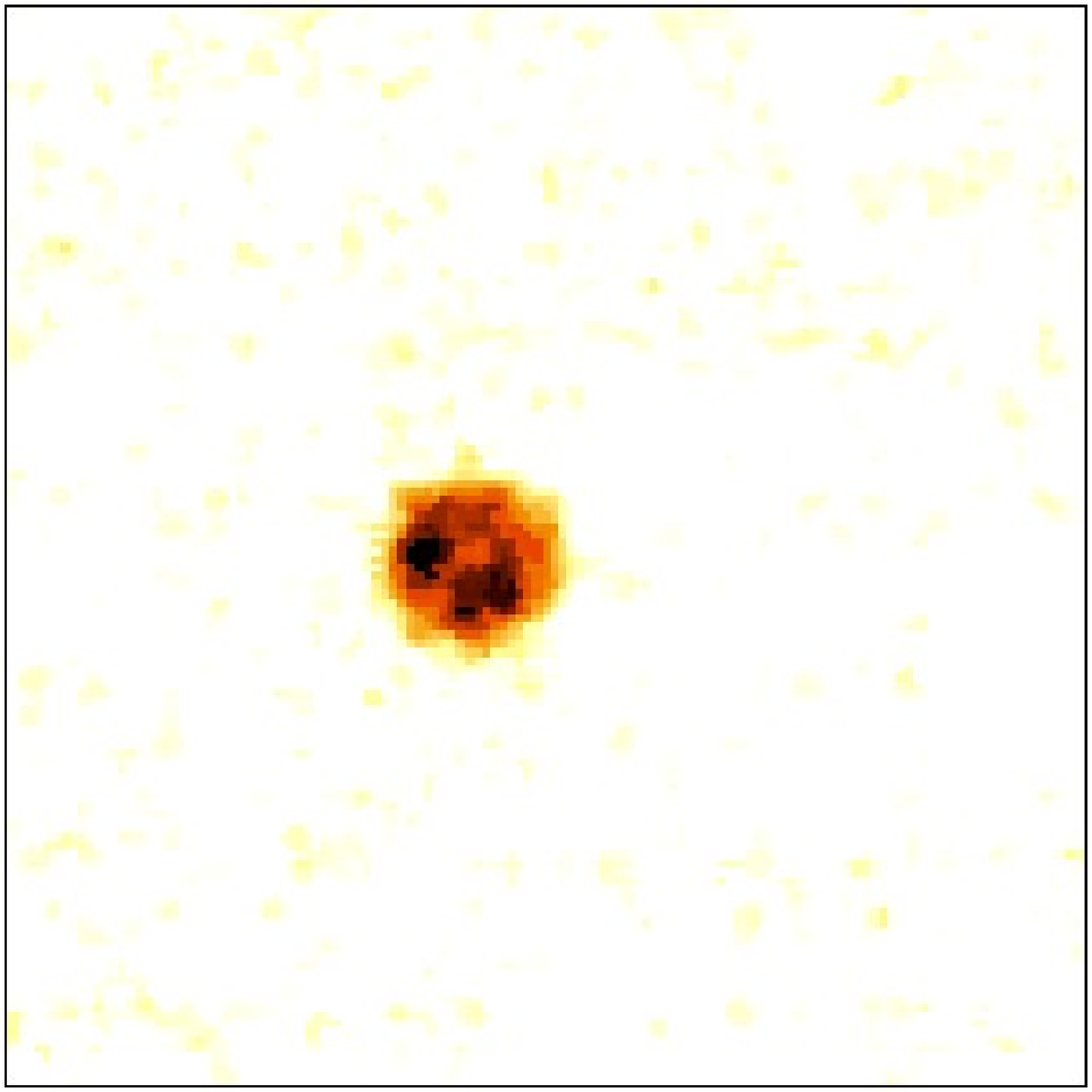}}
\mbox{\epsfclipon\epsfxsize=1.8in\epsfbox[20 20 575 575]{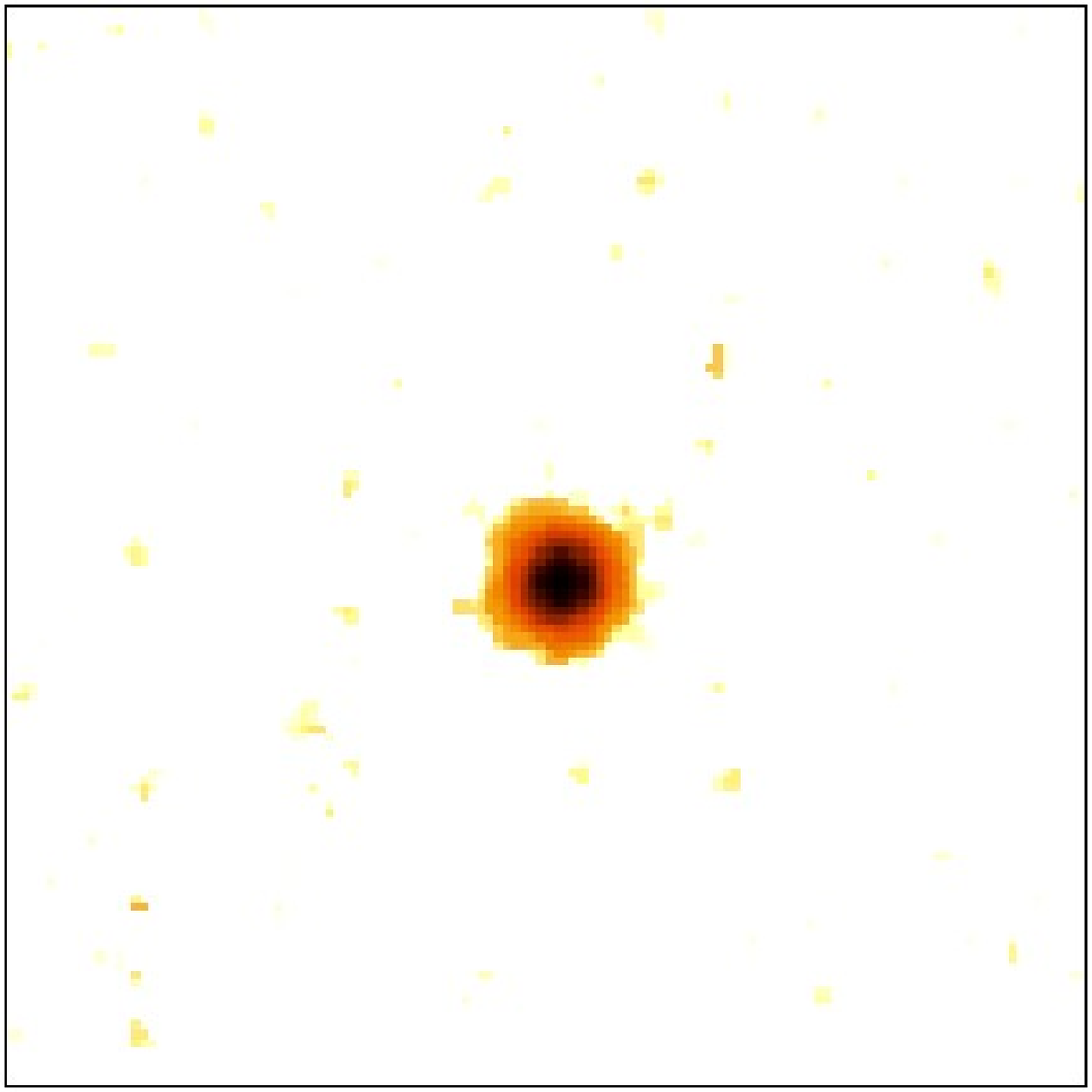}}
\mbox{\epsfclipon\epsfxsize=1.8in\epsfbox[20 20 575 575]{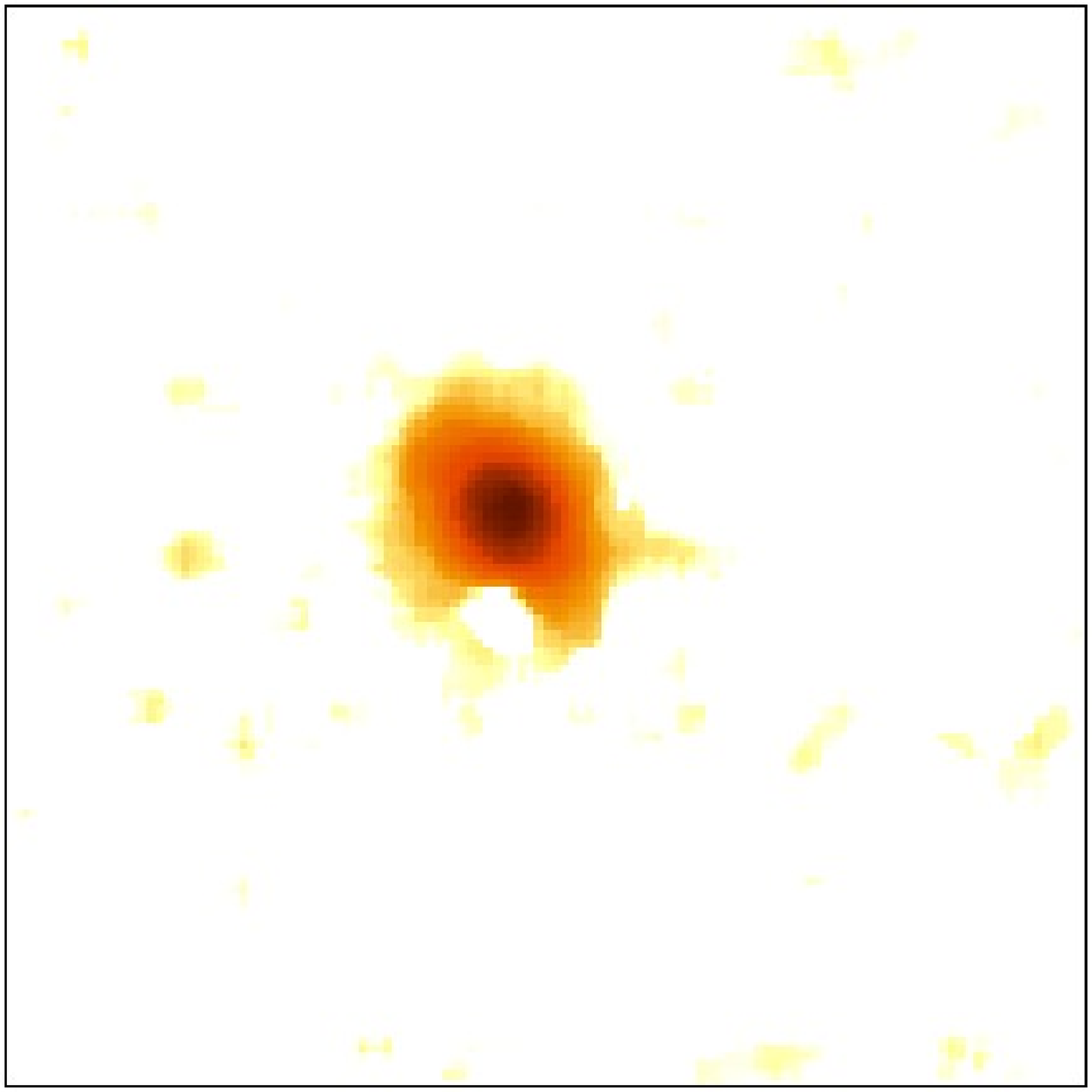}}
\mbox{\epsfclipon\epsfxsize=1.8in\epsfbox[20 20 575 575]{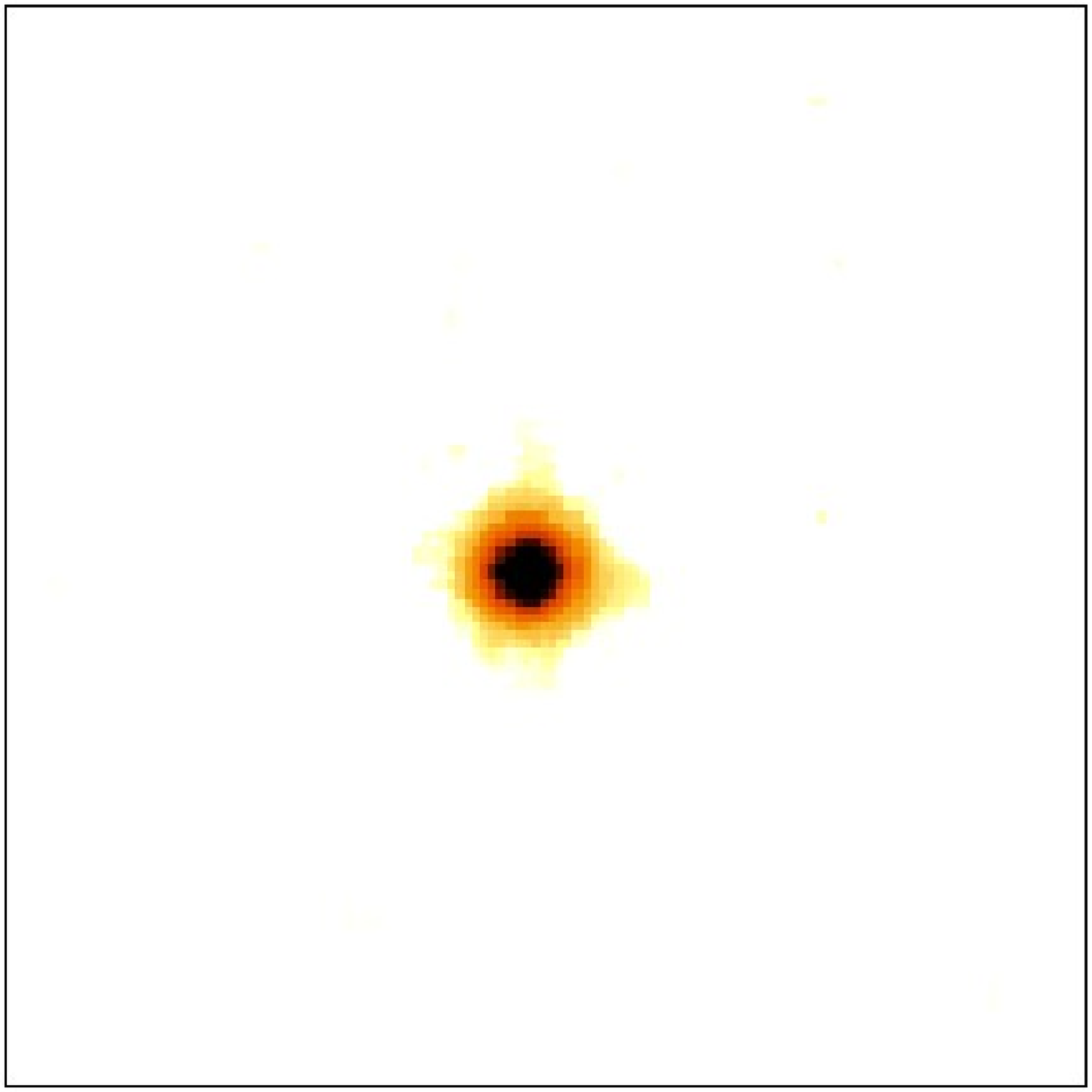}}
\mbox{\epsfclipon\epsfxsize=1.8in\epsfbox[20 20 575 575]{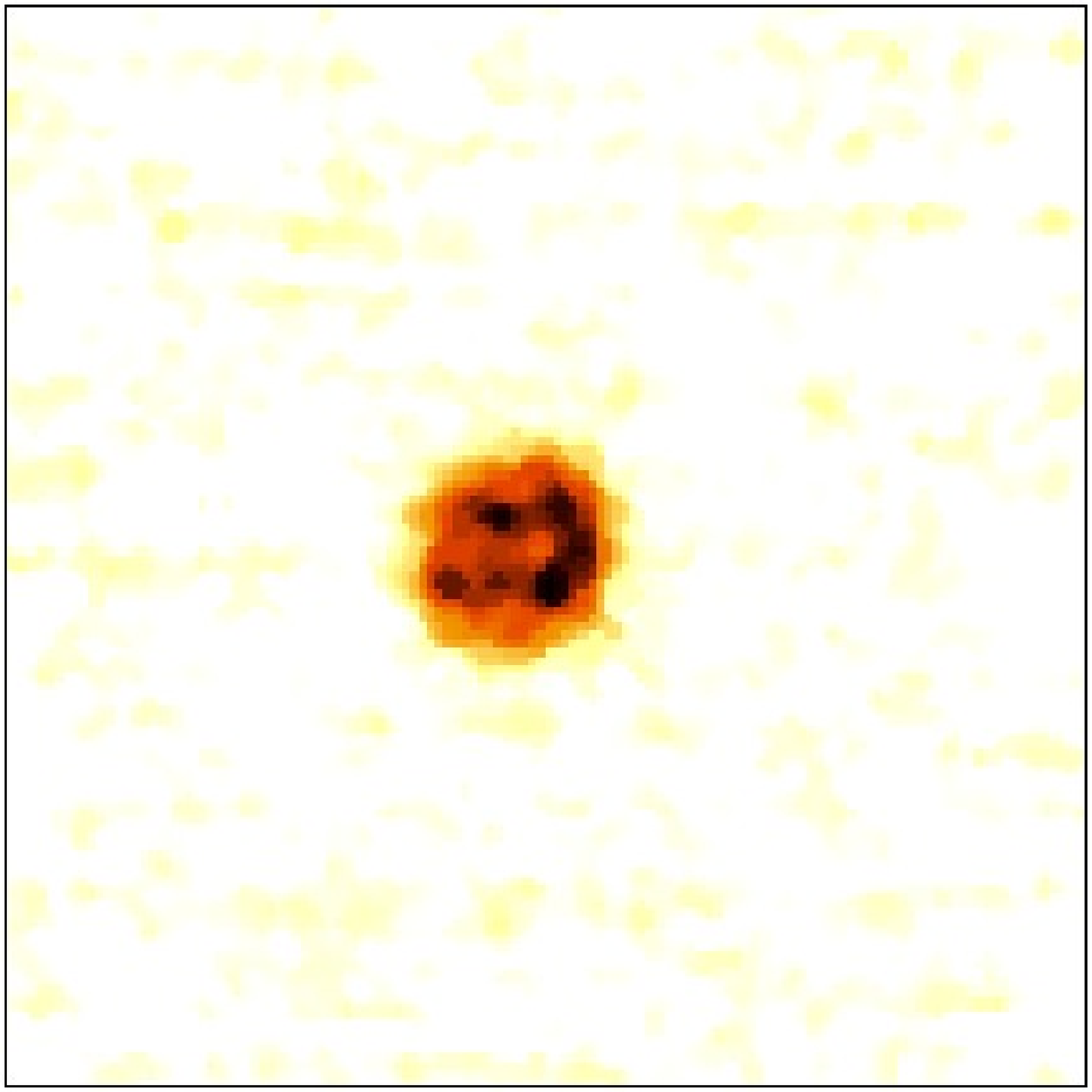}}
\mbox{\epsfclipon\epsfxsize=1.8in\epsfbox[20 20 575 575]{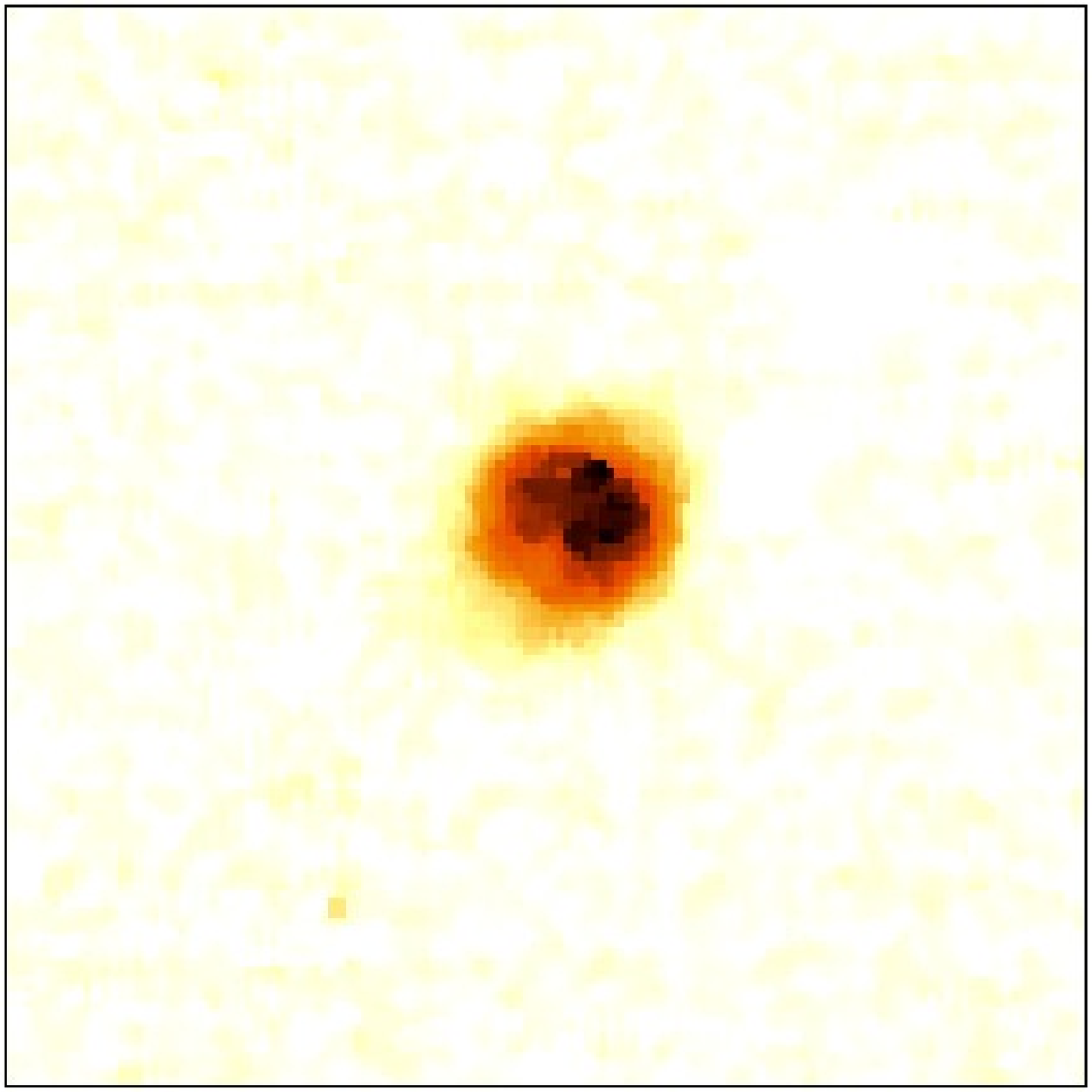}}
\mbox{\epsfclipon\epsfxsize=1.8in\epsfbox[20 20 575 575]{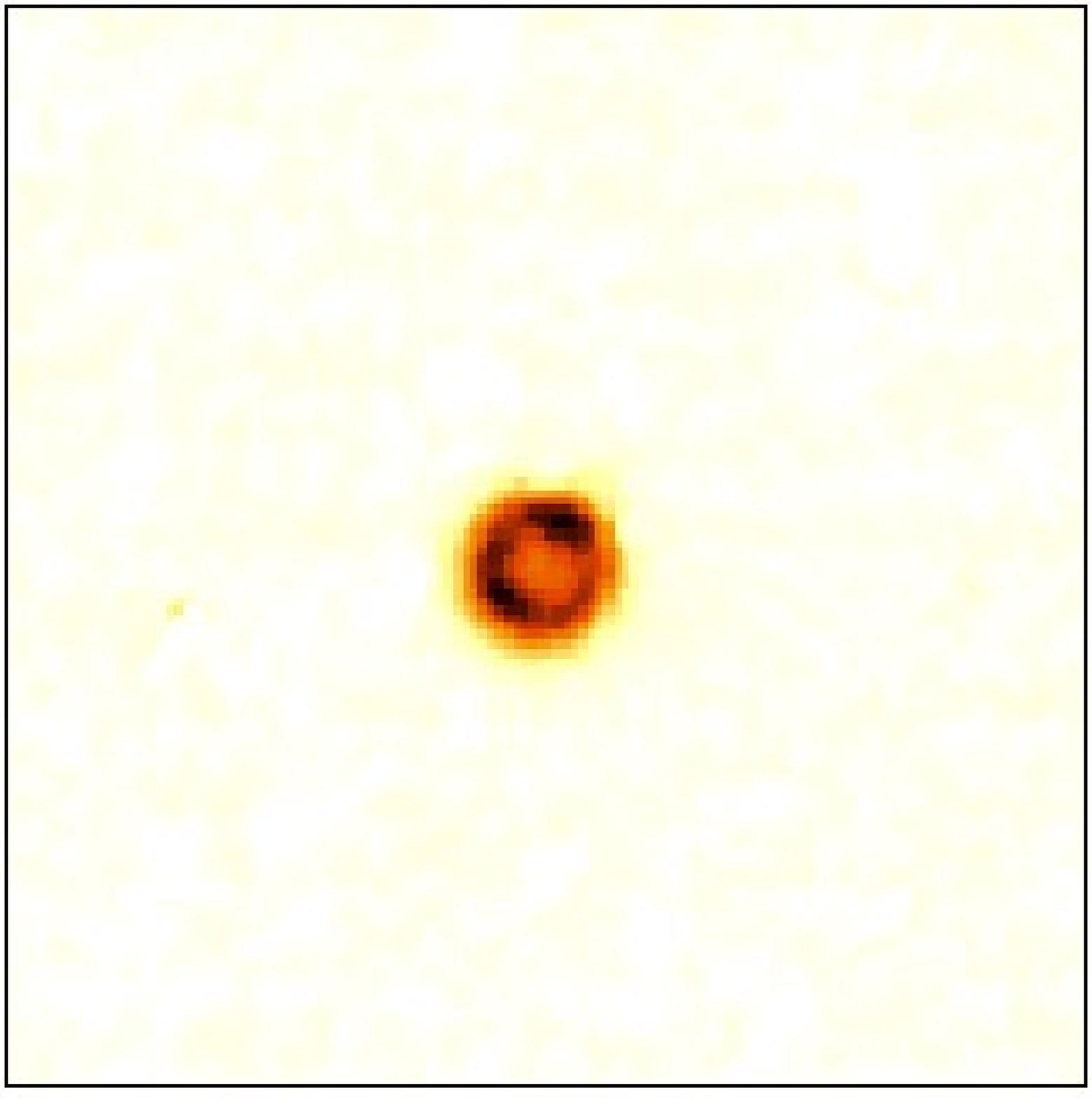}}
\caption{The different morphological PNe classes can be seen in this
figure. In particular PTB14 and PTB20 are round/compact (RC), PTB13
and PTB34 are round/ring (RR) while PTB19 and PTB31 are elliptical (E)
and PTB7 and PTB24 are bipolar (B) with the latter presenting a
quadripolar morphology.} \label{fig1}
\end{figure}

Follow-- up observations (images and spectra) were obtained with the
1.3 m telescope at the same site during 2001--2003. The images were
performed in H$\alpha+$[N {\sc ii}] in order to study the morphology
of the PNe and also measure their angular extent while their
low--resolution spectra confirmed their photoionized nature. All new
PNe can be seen in Paper I and Boumis et al. (2003b; Paper II) while
images of some of the new PNe are shown in Figure~\ref{fig1}, where their
different morphological classes can be seen. The image size is 65
arcsec on each side. North is at the top and east to the left.

\section{Preliminary morphological results}

The H$\alpha+$[N {\sc ii}] images as well as the low resolution
spectra of the newly discovered PNe were used for a more detailed
morphological study. A morphological type was assigned to the new PNe
according to Manchado et al. (2000). In most cases, a spherical
well-defined shell (star-like or ring) was found, elliptical nebulae
are also present while a small number appears to be bipolar. In
particular, 60 percent are round - with 63\%\ of these presenting a
ring--like structure (e.g. PTB34) and the remaining 37\%\ being
compact (e.g. PTB20) - 31\%\ are elliptical (e.g PTB19) and 9\%\
bipolar (e.g PTB7). Note that the morphological class of some PNe
could not be distinguished clearly and even though these have been
classified as round, the possibility that they are elliptical cannot
be ruled out. Also, following the Manchado et al. (2000)
classification, quadripolar PNe (e.g. PTB24) were included in the
bipolar class.  The electron density was found to be different for
each morphological class. The median value for ellipticals is higher
than that for the round and bipolar PNe. Also, a first indication
shows that the [N {\sc ii}]/H$\alpha$~and N/O ratios are higher for
round than for elliptical PNe. However, the abundances in N/O and He
generally showed low N/O and He, implying old progenitor stars. More
analytical results are presented in Papers I, II.

\acknowledgments

Skinakas Observatory is a collaborative project of the
University of Crete, the Foundation for Research and Technology-Hellas
and the Max-Planck-Institut fur Extraterrestrische Physik.

\end{document}